\documentstyle[psfig,prl,aps]{revtex}
\begin{document}
\draft 
\twocolumn[\hsize\textwidth\columnwidth\hsize\csname
@twocolumnfalse\endcsname
\title{Annealing schedule from population dynamics\footnote{tobe}}\footnote{tobe2}    
\author{Stefan Bornholdt\footnote{email}}
\address{
Institut  f\"ur Theoretische Physik, Universit\"at Kiel, 
Leibnizstrasse 15, D-24098 Kiel, Germany \\
bornholdt@theo-physik.uni-kiel.de
}
\date{received 23 June 1998; revised manuscript received 30 November 1998}
\maketitle
\begin{abstract}    
We introduce a dynamical annealing schedule for population-based 
optimization algorithms with mutation. 
On the basis of a statistical mechanics formulation 
of the population dynamics, the mutation rate adapts to a value 
maximizing expected rewards at each time step. Thereby, the mutation rate 
is eliminated as a free parameter from the algorithm.    
\medskip \\
PACS number(s): 05.50.+q, 87.10.+e  07.05.Mh  02.60.Pn
\end{abstract}    
\pacs {To be published in Phys.\ Rev.\ E {\bf 59} (1999).}
]   
\section{Introduction} 
Population-based optimization algorithms \cite{holland,rechenberg,ec} have been   
successfully applied to problems in physics \cite{sg,prb} and beyond \cite{ecappl}. 
This class of algorithms is based on the simultaneous tracking of more than 
one point in search space (a ``population'', in analogy to biological evolution 
\cite{darwin}), in order to make trapping in local optima less likely
during the process of optimization. 
In addition, stochastic noise is used to generate random 
displacements of the search points (``mutations''), performing a local optimization. 
A major problem in these algorithms is the adjustment of the noise level 
for a given optimization task. In the beginning of the search, high noise levels 
help to identify promising regions of the search space, 
while for a subsequent fine tuning low noise works best. 
This problem is well known from simulated annealing \cite{sa}, an optimization
algorithm where noise is introduced by means of a formal temperature. 
Lowering, or ``annealing'', the temperature from high to low values in the course 
of the optimization leads to improved results compared to 
an optimization at fixed temperatures. 
However, there remains the problem of choosing a suitable annealing schedule for 
a given optimization problem \cite{sa}.   
The same problem occurs in population-based optimization algorithms and will be 
addressed in the remainder of this paper. 

For some population-based algorithms heuristics have been
proposed that adjust the noise rate during optimization. 
For example, when noise is implemented as 
random steps of fixed Euclidean distance in the search space, 
its step size can be adjusted according to an estimate for the most 
promising next step on the basis of the previous one \cite{rechenberg}. 
Another approach, taken by Davis \cite{davis}, uses a set of noise operators competing 
for high scores in producing low energy search points.             
Although such approaches work in practice, they have not been formally established
on the basis of a dynamical formulation of the algorithm. 
A major problem is the enormous complexity of the dynamics of population-based algorithms.
    
Recently, this problem has attracted parts of the 
physics community, applying statistical mechanics methods to the algorithm dynamics.   
Pr\"ugel-Bennett and Shapiro \cite{pbs1} described the average population dynamics 
in terms of the distribution of energy values in the population at each time step. 
The observables are the cumulants of the energy distribution of the population. 
Selection of the ``fittest'' (the low energy members) is completely defined in 
terms of these variables. For certain energy functions, this enables a prediction 
of the algorithm dynamics to high accuracy over large numbers of generations \cite{pbs2}. 
In Ref.\ \cite{pbs3} it was proposed to 
use this formalism in order to determine an annealing schedule from the predicted dynamics.  
However, each of the two immediate routes faces a major obstacle: The analytical 
approach is only feasible for exactly known energy functions with simple properties and 
involves a complicated maximum entropy calculation. The alternative way via measuring the 
current cumulants during evolution is spoiled by large sample-to-sample fluctuations. 
Here we propose 
a model which, though inspired by this, does not have to deal with these problems. 
For an {\em a priori} unknown energy function (being the 
usual case when dealing with optimization problems) a less formal framework
is needed. This is supplied by a dynamical model based on energy 
correlation formulated in Ref.\ \cite{sb}. 
It will be used here to predict an optimal noise rate by maximizing the expected 
performance of the algorithm in each time step. 
In the following we will define two test functions and two algorithms to be considered. 
A model for the prediction of the mutation effects will then be given. 
An improved algorithm will be defined on the basis of this model. 
It is then applied to the test functions and its performance compared to  
standard versions of the respective algorithms.   

\section{Test System}  
Let us consider an optimization problem in terms of the minimization 
of a real valued function $E(S)$ on a binary search space. Its value is the 
energy of the test point $S$ which is to be minimized, equivalent to a search for 
the ground state energy of a physical system \cite{footnote}.   
In biological terms this corresponds to the negative ``fitness'' 
of an organism to be maximized for survival. 
The discrete search space will be parametrized through the binary representation 
$S \in \{\pm1\}^N$ of length $N$. 

Two functions serve as examples: 
one purely additive energy function, and another with many local optima.  
The first problem is a random field paramagnet
\begin{eqnarray}
E_\alpha = \sum\limits_{i=1}^N J_i\; S_i^\alpha + \kappa_1^0
\label{rpmod}
\end{eqnarray}
with random couplings $J_i$ taken from a Gaussian distribution
with mean $0$ and variance $1$. The $N$ spins $S_i^\alpha$ with
$i=1,\dots,N$ and $S_i^\alpha = \pm 1$ form the genetic string
of the member $\alpha$ of the population.
The second function is the NK model energy function \cite{nkmodel}
as an example for a hard search problem with many local minima. It is defined through 
\begin{eqnarray}
E_\alpha = \sum\limits_{i=1}^N E_i(S_i^\alpha;S_{i_1}^\alpha,
           \dots , S_{i_K}^\alpha)
\label{nkfunc}
\end{eqnarray}
with $2^{K+1}$ random energy values $E_i(S^\alpha)$ drawn from a uniform
distribution over the interval $[0,1]$ and a randomly chosen permutation of
sites $i_1$ to $i_K$, both for each $i$.
Originally, this function has been formulated for the study of evolution on tunably
rugged energy landscapes with application to the evolution of the immune response
\cite{immune}.
These functions will be minimized by means of a population-based algorithm 
which is defined as follows:
First, a random ensemble of search points $S_\alpha$ with $\alpha =1,\dots,P$ 
and energies $E_\alpha = E(S_\alpha)$ is chosen (a ``population'').  
One time step of the algorithm consists of the following procedure: 
\begin{enumerate}
\item Select the member with the lowest energy.         
\item Reproduce it once.         
\item Replace the member with the highest energy by the new copy.  
\item ``Mutate'' all members except the original 
one with the lowest energy by inverting 
each spin with a small fixed probability $\gamma$.  
\end{enumerate}
Repeating these steps then forms an evolutionary algorithm searching for 
low lying energy states. It is driven by selection lowering the mean energy 
of the population and mutation increasing the variance.    

For comparison let us also look at a simplified algorithm, 
a stochastic gradient descent. 
After the initial population is chosen as above, the following steps are taken: 
\begin{enumerate}
\item Select the member with the lowest energy.         
\item Reproduce it $P-1$ times.      
\item ``Mutate'' the new copies by inverting each spin with a small fixed probability $\gamma$.  
\item Replace all members, except the one with the lowest energy, 
by the mutated copies. 
\end{enumerate}
Here, the offspring of the best member takes over the entire population in each time step. 

\section{Modeling the algorithm dynamics}  
We will model the dynamics of these algorithms in terms of the energy distribution $\rho(E)$ 
of the population expressed as an expansion in cumulants. 
The energy distribution $\rho(E)$ of a population is the natural 
quantity for the selection operator, which solely acts on the 
energy values of the search points. The expansion in cumulants of 
the energy distribution $\rho(E)$ has been shown to be a useful 
approximation for population-based algorithms \cite{pbs2}.  
At each time step, the evolving population is then approximated by 
a set of these variables. 
When also modeling the mutation operator, one has to be 
more careful. Mutation, acting on the underlying 
representation instead of the energy values themselves, 
requires additional assumptions to model its dynamics in terms of $\rho(E)$. 
For example, one could obtain a maximum likelihood estimation for 
the underlying spin states corresponding to a given energy 
cumulant, and use this to calculate the expected effect of the 
mutation operator. However, such a procedure requires simple 
energy functions to allow for the calculation, and, of course, 
a complete knowledge of the energy function, which is usually not 
available for realistic optimization problems.

Here we use a different approach which is based solely on a 
phenomenological parameter that is accessible by measurement 
if the energy function is not known \cite{sb}. 
In particular, we will use a model for the 
lowest order average dynamics of the mutation operator on the basis of 
the energy correlation $m$ of mutation on a given energy landscape $E(S)$,  
\begin{eqnarray}
m = \frac{
        \left< E_\alpha E_\alpha^m \right>_{\alpha, \mbox{\scriptsize mut}}
        - \left< E_\alpha \right>_\alpha \left< E_\alpha^m \right>_{\alpha, \mbox{\scriptsize mut}}
    } {
        \left< E_\alpha^2 \right>_{\alpha} - \left< E_\alpha \right>_\alpha^2
    }, 
\label{mdef}
\end{eqnarray}
where $\left< \;\;\; \right>_\alpha$ denotes an average over the 
population, and $\left< \;\;\; \right>_{\mbox{\scriptsize mut}}$ 
an average over all possible mutation events. 
The energy correlation $m$ is a measure of how strongly, on 
average, the energy of a mutant is correlated to that of its parent, 
for a given mutation operator applied to a given energy function. 
Such correlations form the backbone on which the search 
process in mutation-based algorithms proceeds. 
Energy correlations can be measured for many hard optimization 
problems, as were recently classified in Ref.\ \cite{stadler}.  
In this framework, the energy distribution of a population
after mutation can be approximated by its cumulants 
as a function of $m$,  
\begin{eqnarray}
\kappa_1^m &=& m \; \kappa_1  + (1 - m) \; \kappa_1^0
\nonumber \\
\kappa_2^m &=& m^2 \; \kappa_2  + (1 - m^2) \; \kappa_2^0
\nonumber \\
\kappa_3^m &=& m^3 \; \kappa_3
\nonumber \\
\kappa_4^m &=& m^4 \; \kappa_4,   
\label{cum}
\end{eqnarray}
where $\kappa_1^0$ and $\kappa_2^0$ are the energy mean and variance of a random 
initial population. This model was derived in Ref.\ \cite{sb}.  
The underlying assumption of this model is that the population of 
the algorithm (not the landscape itself) can be expanded 
in cumulants around a Gaussian. In fact, one observes that the initial
random population for many real optimization problems fulfills this 
requirement well.   
In the framework of this expansion, the above model uses $m$ 
to predict the expected energy distribution of the population in the next time step. 
Such a model, based on energy correlations, further assumes 
that the lowest order correlation $m$ contains major information 
about the average effect of the mutation operator. 
It has been shown to be useful to describe 
the dynamics of a population-based algorithm over at least $200$   
generations, both for correlated and poorly correlated landscapes \cite{sb}. 

How can such a model be used to improve an optimization algorithm? 
Let us look at a numerical example for the dynamics of a stochastic gradient descent under 
a fixed mutation rate $\gamma$, as shown in Fig.\ \ref{rpada}.  
Optimization of the first test function (\ref{rpmod}) is shown with a 
stochastic gradient descent, searching for the minimal energy 
configuration of a random paramagnet of $N=128$ spins in an external field. 
For a large mutation rate $\gamma$ one sees that  
the early gain is large, whereas for small $\gamma$, as shown by the solid curve, 
a poor early gain is balanced later by a slow but steady improvement.        
For optimization problems involving computationally costly energy evaluation, 
this behavior poses a severe problem. 
Knowledge of the latter stages of the dynamics would be needed 
at the beginning in order to be able to choose an optimal 
$\gamma$. 
In the following, this problem will be addressed 
through a variable mutation rate $\gamma(t)$,  
that combines the advantages of both regimes of the mutation rate $\gamma$. 

\section{Annealing the mutation rate} 
For this purpose, the expected best member of a population after mutation 
$\left< E_{\mbox{\scriptsize min}} \right>$  
is evaluated on the basis of the energy distribution 
$\rho^m(E)$ of the population after mutation given in terms of cumulants
$\kappa_i^m$. The expectation value for the lowest energy occurring in a set
of $P$ samples \cite{magnus} drawn from the post-mutation distribution $\rho^m(E)$ is  
\begin{equation}   
\left< E_{\mbox{\scriptsize min}} \right> = P \; 
\int\limits_{-\infty}^\infty dE_1\; E_1 \; \rho^m(E_1) \; 
\prod\limits_{n=2}^P  \; 
\int\limits_{E_1}^\infty \; dE_n \; \rho^m(E_n).      
\end{equation}   
In the Gaussian approximation a saddle point expansion yields, to leading order,    
\begin{equation}   
\left< E_{\mbox{\scriptsize min}} \right> =    
\kappa^m_1 - \sqrt{2 \kappa^m_2\ln P}.   
\label{estimate}   
\end{equation}   
Inserting the post-mutation distribution (\ref{cum}), the expected energy of the best member 
after mutation $\left< E_{\mbox{\scriptsize min}} \right>$ can be minimized in terms of $m$. 
The resulting mutation correlation $m_{\mbox{\scriptsize opt}}$
is then used to choose the mutation rate $\gamma$ in the forthcoming 
mutation step, thereby optimizing the expected best member of the next generation. 
Unfortunately, this method is plagued by large 
fluctuations in the measured moments of the energy distribution. 

Therefore, let us first look at the expected dynamics of the stochastic gradient descent 
where this problem does not occur. 
Following Eq.\ (\ref{cum}), the energy distribution after the mutation step 
is given in the Gaussian approximation by  
\begin{eqnarray}
\kappa^m_1 &=& m \; E_{\mbox{\scriptsize min}}(t)  + (1 - m) \; \kappa_1^0
\nonumber \\
\kappa^m_2 &=& (1 - m^2) \; \kappa_2^0.  
\label{7}
\end{eqnarray}
Inserting this into Eq.\ (\ref{estimate}), and minimizing     
the expected best member of the next generation $\left< E_{\mbox{\scriptsize min}}(t+1) \right>$ 
with respect to $m$, yields an estimate for an optimal correlation:  
\begin{equation}   
m_{\mbox{\scriptsize opt}}  = \sqrt{\frac{\left(E_{\mbox{\scriptsize min}}(t)-\kappa_1^0\right)^2}{2 
         \; \kappa_2^0 \; \ln(P) 
         + \left(E_{\mbox{\scriptsize min}}(t)-\kappa_1^0\right)^2}}.       
\label{mopt}
\end{equation}   
This is subsequently translated into an optimal mutation rate $\gamma_{\mbox{\scriptsize opt}}$ via   
\begin{eqnarray}
m = 1-2\gamma,    
\label{mrp}
\end{eqnarray}
which is derived from the known energy function. 
Each time step of the modified algorithm can now be described as follows: 
\begin{enumerate}
\item Determine the lowest energy $E_{\mbox{\scriptsize min}}$ in the population. 
\item Calculate the optimal correlation $m_{\mbox{\scriptsize opt}}$ from Eq.\ (\ref{mopt}) 
and calculate the mutation rate $\gamma_{\mbox{\scriptsize opt}}$ from this. 
\item Select the member with the lowest energy.         
\item Reproduce it $P-1$ times.      
\item ``Mutate'' the new copies by inverting each spin with the 
mutation rate $\gamma_{\mbox{\scriptsize opt}}$.
\item Replace all members, except the one with the lowest energy, 
by the mutated copies. 
\end{enumerate}
Starting from an initial condition as above and iterating this step results in an algorithm 
with an adaptive mutation rate. How it applies to the above test function is 
shown in Fig.\ \ref{rpada}. 
At each time scale, the evolution of the lowest energy member of the evolving population
compares well to the respective best ``fixed mutation rate algorithm''.    
No explicit knowledge of favorable ranges of the mutation rate $\gamma$ is used, 
thus removing the free parameter $\gamma$ from the algorithm.  
Applying the formalism to the NK-model function (\ref{nkfunc}), and using the relation 
between parent child correlation $m$ and mutation rate $\gamma$, derived as 
\begin{eqnarray}
m = (1-\gamma)^{K+1},   
\label{mnk}
\end{eqnarray}
a comparable result is obtained (Fig.\ \ref{nkada}).  

A similar procedure can also be carried out for 
the full population-based algorithm with sparse replication. 
Again, Eq.\ (\ref{cum}) is used to adjust the mutation rate in the next generation
to a value that maximizes the expected gain. 
In order to avoid large fluctuations, which would be incompatible with a smooth evolution, 
we do not base the prediction on the cumulants of the current energy distribution in the 
population, but rather on $E_{\mbox{\scriptsize min}}$ alone. 
This is done in the spirit of Eq.\ (\ref{7}), which is less likely to fluctuate than 
the prediction based on the full cumulants $\kappa_i$.   
However, $E_{\mbox{\scriptsize min}}$ still relates to the dynamics of a mixed population, 
and proves to be useful in modeling the population dynamics under mutation. 
Depending on the mutation stregth,  
a number of former mutants are still correlated with the new offspring, 
in addition to the one copy of $E_{\mbox{\scriptsize min}}$ made per generation.  
Let us assume that a number of $M$ members of the population 
are strongly correlated with the new offspring. 
For simplicity we further assume that the remaining members are completely 
uncorrelated and treat them as random. 
In this approximation, the integral for the expected best member of a population
can be written as  
\begin{eqnarray}   
&&
\left< E_{\mbox{\scriptsize min}} \right> 
= M \; \int\limits_{-\infty}^\infty dE_1\; E_1 \; \rho^m(E_1) 
\cdot 
\nonumber \\  
&&\cdot 
\left[ \int\limits_{E_1}^\infty \; dE_2 \; \rho^m(E_2) \right]^{M-1}
\; \left[ \int\limits_{E_1}^\infty \; dE_3 \; \rho^0(E_3) \right]^{P-1-M}
\nonumber \\  
&&+ (P-1-M) \; \int\limits_{-\infty}^\infty dE_1\; E_1 \; \rho^0(E_1) 
\cdot 
\nonumber \\  
&&\cdot 
\left[ \int\limits_{E_1}^\infty \; dE_2 \; \rho^m(E_2) \right]^{M}  
\; \left[ \int\limits_{E_1}^\infty \; dE_3 \; \rho^0(E_3) \right]^{P-2-M}.      
\end{eqnarray}   
It is solved using a saddle point expansion in the Gaussian approximation, 
considering the limit where the distributions $\rho^m$ and $\rho^0$ 
move sufficiently apart from each other 
(due to $E_{\mbox{\scriptsize min}}$ moving away from the random 
population distribution),     
where one can neglect their mutual variations. 
One obtains 
\begin{eqnarray}   
\left< E_{\mbox{\scriptsize min}} \right> &=&    
\kappa^m_1 - \sqrt{2 \kappa^m_2\ln (M-1)} 
\nonumber \\  
&+& \kappa^0_1 - \sqrt{2 \kappa^0_2\ln (P-2-M)}.   
\label{estimate2}   
\end{eqnarray}   
The expected $\left< E_{\mbox{\scriptsize min}}(t+1) \right>$ 
of the next generation based on Eq.\ (\ref{7}) is 
then minimized by the mutation rate 
\begin{equation} 
m_{\mbox{\scriptsize opt}} = \sqrt{ \frac{ (E_{\mbox{\scriptsize min}}(t) - \kappa_1^0)^2}
{2\kappa_2^0\ln(M-1) + (E_{\mbox{\scriptsize min}}(t) - \kappa_1^0)^2}}.  
\label{14} 
\end{equation}   
Finally, the number of correlated members $M$ 
in the population remains to be specified.   
For a lowest order estimate 
let us consider a member with energy $E_{\mbox{\scriptsize min}}$ 
and mutate it $k$ times. We then require that its energy does not, 
on average, move away more than $\sqrt{2\kappa_2^m}$ 
from the current value of $E_{\mbox{\scriptsize min}}$, i.e., 
\begin{equation}
E_{\mbox{\scriptsize min}}  + \sqrt{2\kappa_2^m}
\; > \; m^k \;  E_{\mbox{\scriptsize min}}  + 
(1-m^k) \; \kappa_1^0.  
\end{equation}  
The exact limit for the number of subsequent 
mutations $k$ depends on the current details of the 
energy values in the population. However, when using Eq.\ (\ref{14})  
as an estimate for the current value of $m$, the energy value
of a mutant decorrelates after only a few mutation steps. 
Therefore, ln$(M-1)$ is estimated to be of the order of $1$ and 
we determine the optimal mutation rate in the algorithm using  
\begin{equation} 
m_{\mbox{\scriptsize opt}} = \sqrt{ \frac{ (E_{\mbox{\scriptsize min}}(t) - \kappa_1^0)^2}
{2\kappa_2^0 + (E_{\mbox{\scriptsize min}}(t) - \kappa_1^0)^2}}.  
\label{16} 
\end{equation}   
This expression is now used for annealing the mutation rate in 
the population based algorithm. The modified time step of the algorithm is defined 
by the following procedure: 
\begin{enumerate}
\item Determine the lowest energy $E_{\mbox{\scriptsize min}}$ in the population. 
\item Calculate the optimal correlation $m_{\mbox{\scriptsize opt}}$ from Eq.\ (\ref{16}),  
and calculate the mutation rate $\gamma_{\mbox{\scriptsize opt}}$ from it. 
\item Select the member with the lowest energy.         
\item Reproduce it once.         
\item Replace the member with the highest energy by the new copy.  
\item ``Mutate'' all members except the original 
one with the lowest energy by inverting 
each spin with the probability $\gamma_{\mbox{\scriptsize opt}}$.  
\end{enumerate}
Again starting from an initial condition as above and iterating this step results 
in an algorithm with annealed mutation rate. 
In Fig.\ \ref{rppop} the evolution of the best population member on the basis of this 
algorithm is compared to runs with fixed mutation 
rates. 
The algorithm adjusting the mutation rate 
compares well to the fixed mutation rate cases at each stage of evolution.  
In Fig.\ \ref{nkpop} the algorithm is applied to the  
NK-model function with similar results. 
For any given resource of CPU time, 
one reaches a level of performance comparable to an optimum fixed mutation rate 
(at the given total evolution time). This is helpful in optimization when the relationship between 
mutation rate $\gamma$ and the algorithm dynamics at later times 
is {\em a priori} unknown. 

\section{Discussion} 
For both algorithms considered above, we have seen how annealing the mutation rate
can be based on a simple dynamical model based on the energy correlation of 
the mutation operator. In the presented examples, 
functions with known analytical properties have been considered, enabling  
a direct calculation of the mutation correlation $m(\gamma)$. 
However, when applying the above method to general optimization problems, 
this functional dependence remains to be established.
For many realistic optimization problems it is well approximated by 
a monotonic function with a simple decay law in the small $\gamma$ regime, 
as classified in Ref.\ \cite{stadler} for a number of different 
optimization problems. For many problems it can be modeled 
using the simple linear approximation $\gamma(m) = 1 - x \; m$. 
In order to apply the above algorithms to optimization problems 
where the energy function is not known, a heuristics that 
measures this relation for a given problem has been defined. 
One possibility is to measure $m$ and improve the estimate for $x$ during 
the run of the above algorithms. This procedure can be defined as follows: 
\begin{enumerate} 
\item Start from an initial estimate for $x$.
\item Measure the mutation correlation $m$ during each time step
of the algorithm using (\ref{mdef}).  
\item Use the measured $m$ to improve the estimate for $x$ in the linear approximation 
(taken as the average over all measured values of $x$ so far). 
\end{enumerate} 
This allows one to apply the method to energy functions 
with no {\em a priori} knowledge of their correlation structure. 
This method has been sucessfully tested using the two energy functions of this study.  

Several extensions remain to be studied, e.g., algorithms where recombination,
or ``crossover'', is present.   
In such algorithms, the annealed mutation as described here is expected to work 
equally well as long as the mutation step does not strongly interact with the crossover. 
Whether the recombination strength can be adapted in a similar way is an open question. 
Another free parameter is introduced by selection, namely, selection strength.
Here a one parameter model exists \cite{pbs1}, and an adaptive adjustment could be discussed as well. 

To summarize, we proposed a mechanism for annealing the mutation rate 
in population-based algorithms. It is based on a statistical mechanics model of the 
population dynamics and a correlation measure of the mutation operator. 
The mutation rate $\gamma$ thereby drops out as a free parameter of the algorithm.   
\medskip \\
\begin{center} 
{\small \bf ACKNOWLEDGMENTS}  
\end{center} 
The author thanks J.L.\ Shapiro for a useful discussion on this subject,   
and, for further stimulating discussions, M.\  Mitchell and R.G.\ Palmer.   
The hospitality and support of the Santa Fe Institute as well as the financial 
support of the Deutsche Forschungsgemeinschaft is greatly acknowledged.

\begin{center} 
{\bf FIGURES}  
\medskip 
\end{center} 

\begin{figure}[htb]
\caption{The evolution of the member with maximum fitness $f=-E_{\mbox{\scriptsize min}}$
is shown at different fixed mutation rates $\gamma$
of a stochastic gradient descent for the random paramagnet.
In comparison, the points show the dynamics of the adaptive mutation algorithm.
In all simulations a quenched average over $200$ runs is shown,
with a random energy function chosen once.
The dotted line denotes the global optimum of the function.
}
\label{rpada}
\end{figure}

\begin{figure}[htb]
\caption{Same as Fig.\ 1 for an evolution on the rugged landscape of
the NK-model energy function with $K=8$.}
\label{nkada}
\end{figure}

\begin{figure}[htb]
\caption{
Adaptive mutation in the population-based algorithm
compared to the fixed mutation case for a random paramagnet,
with conventions chosen as in the previous figures.
The dotted line denotes the global optimum of the function.}
\label{rppop}
\end{figure}

\begin{figure}[htb]
\caption{Adaptive mutation in the population-based algorithm
compared to a fixed mutation rate for the NK model.}
\label{nkpop}
\end{figure}


\begin{thebibliography}{99}
\bibitem{footnote}
In general, one has to be careful to carry  
the analogy further to a physical system, since the dynamics does not have to follow all 
major principles of known dynamics in a physical system. That is, detailed balance may 
well be violated in optimization, while ergodicity remains essential.
\end{thebibliography}
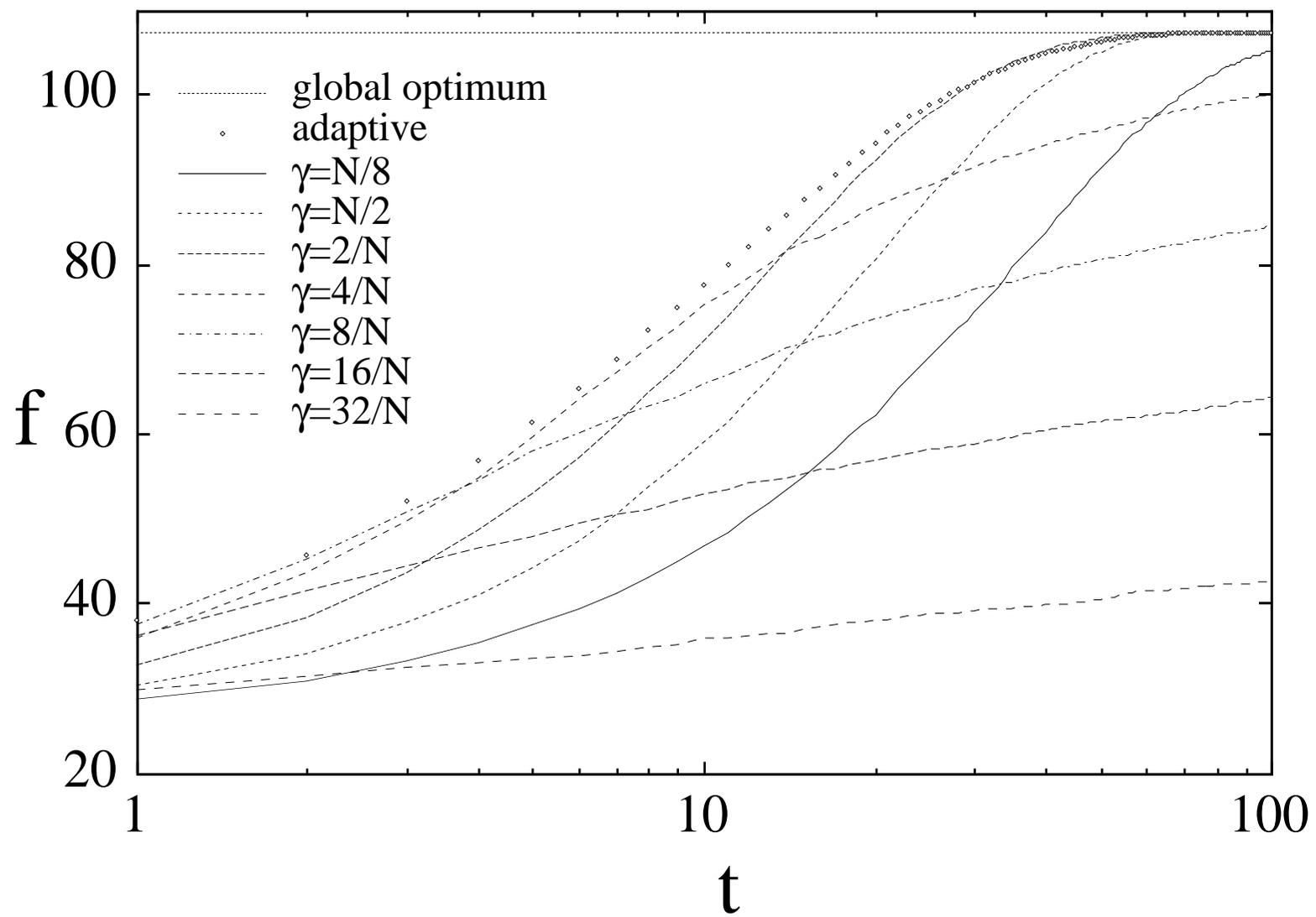
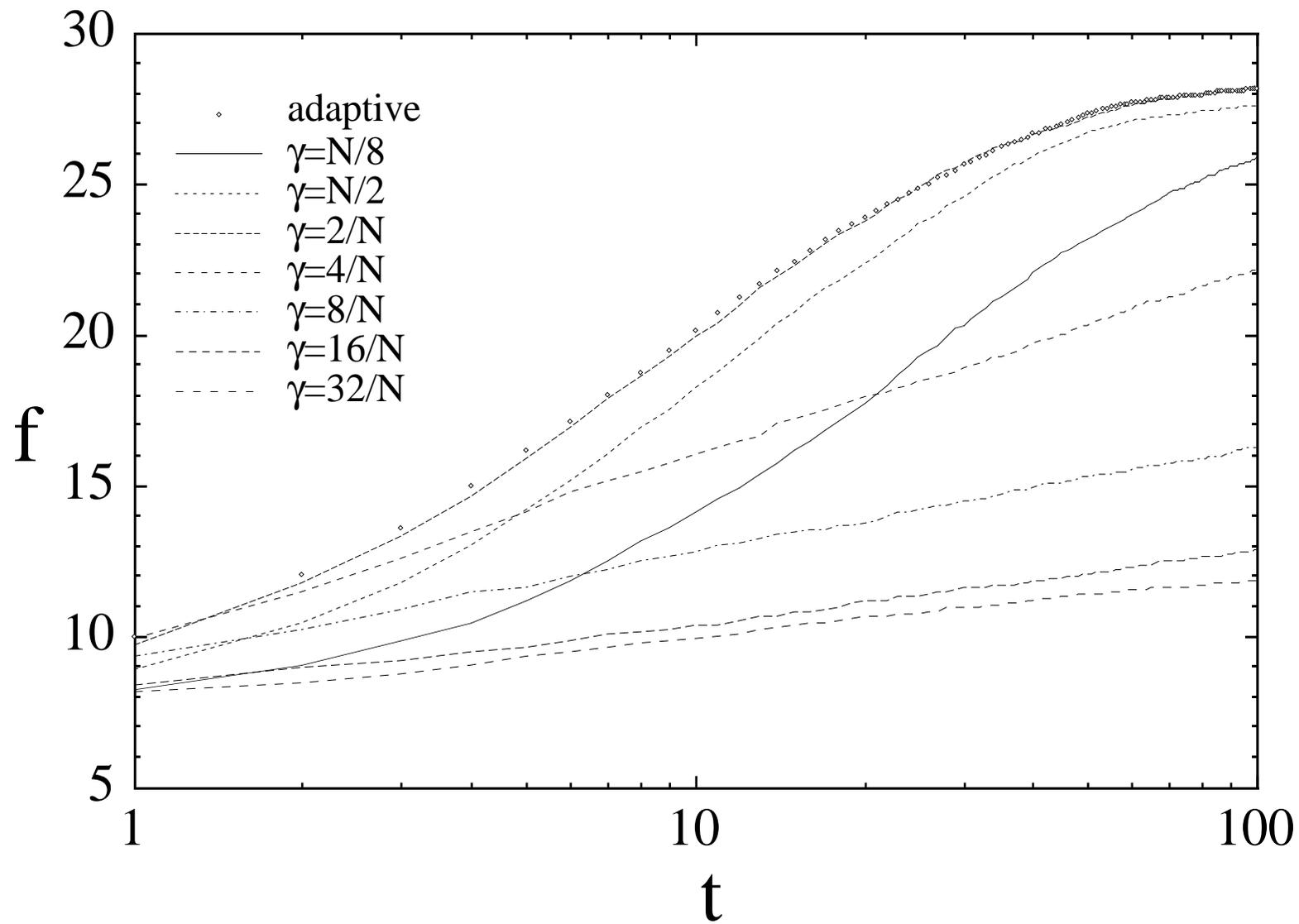
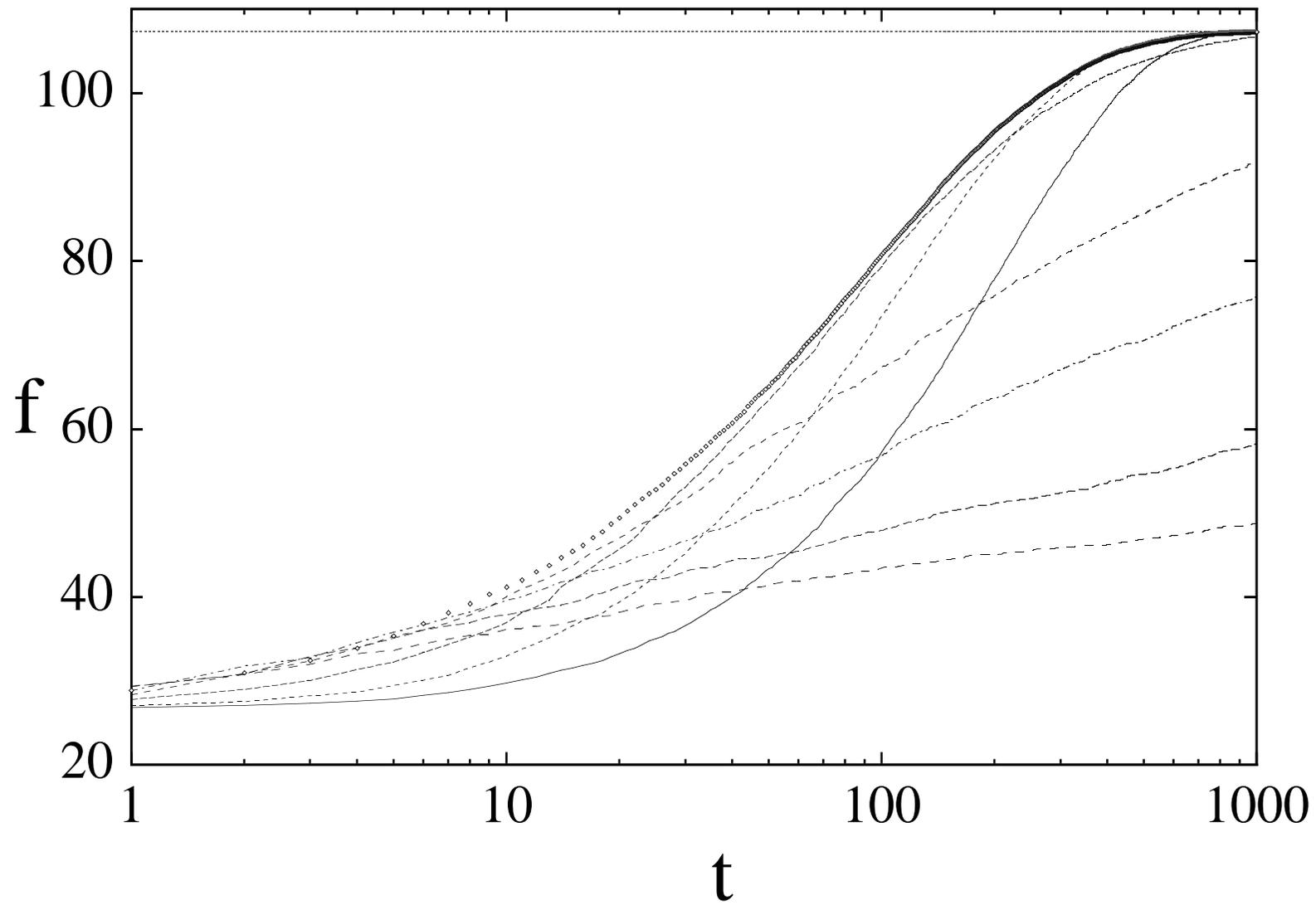
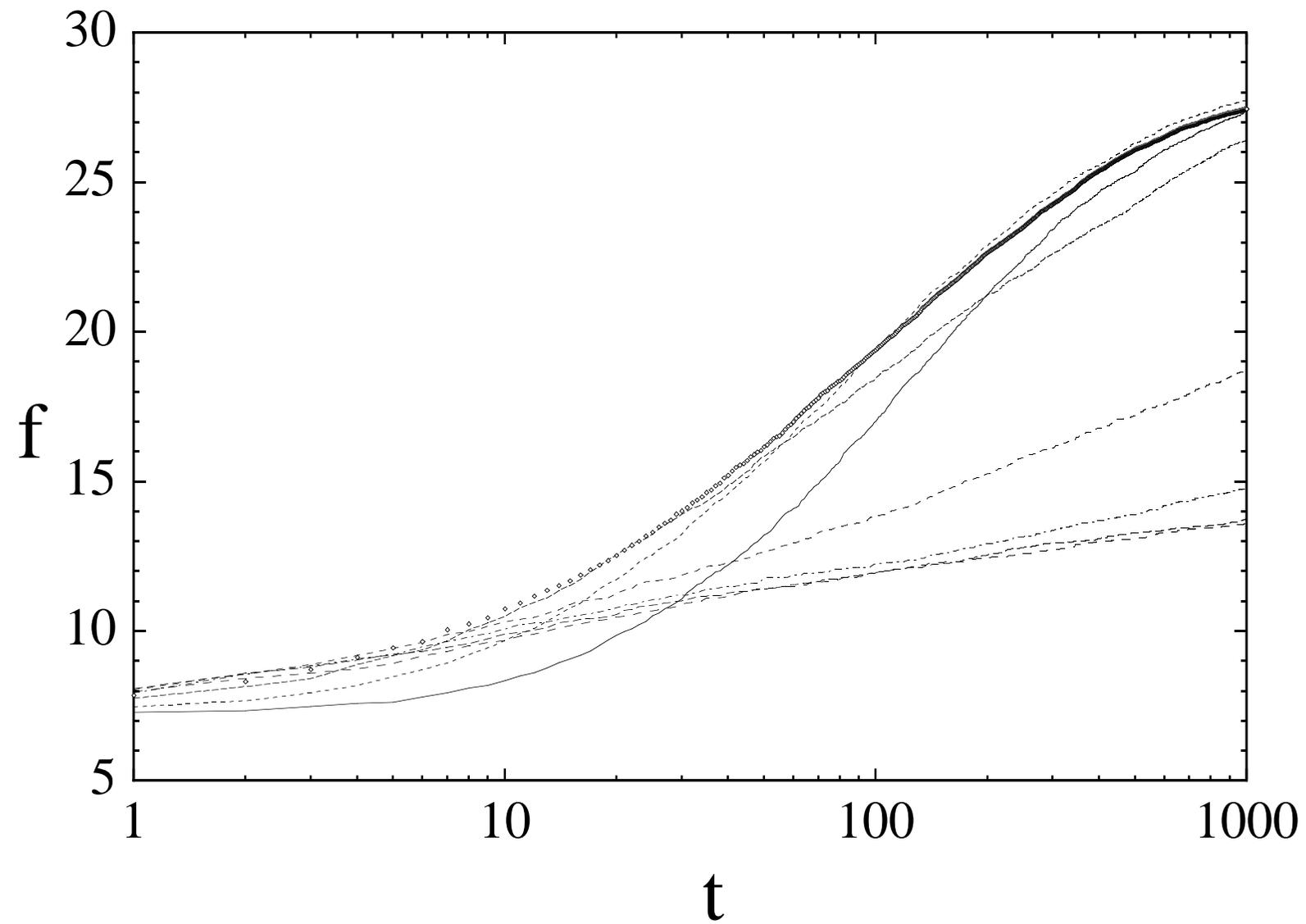
\end{document}